\documentclass{PoS}
\PoS{PoS(LAT2005)236}
\title{Tuning anisotropies for dynamical gauge configurations}
\ShortTitle{Tuning Anisotropies for dynamical gauge configurations}
\author{\speaker{Richie Morrin}, Mike  Peardon, Sin\'ead M. Ryan \\
	School of Mathematics, 
	Trinity College, Dublin, Ireland\\
	E-mail: \email{rmorrin@maths.tcd.ie},
		\email{mjp@maths.tcd.ie},
		\email{ryan@maths.tcd.ie}}
\abstract{We present methods and results for the tuning of quark and gluon anisotropies for improved actions in $N_f = 2$ QCD.}
\FullConference{XXIII International Symposium on Lattice Field Theory\\
		25-30 July 2005\\
		Trinity College, Dublin, Ireland}

\begin{document}

\section{Introduction}
The use of anisotropic lattices in QCD allows for improved resolution in correlation function decays of heavy particles with a minimal increase in computational workload. The anisotropic lattice formulation has also proven to be particularly advantageous in finite temperature QCD \cite{jonivar}. However, the use of an anisotropic lattice introduces an extra parameter, $\xi = \frac{a_s}{a_t}$ ($a_s$ and $a_t$ being the lattice spacings in the spatial and the temporal directions respectively), into our calculations. As with other parameters the bare input values for $\xi_g$ (in the gluon action) and $\xi_q$ (in the quark action) must be tuned correctly in order that a measurement of the output, renormalised, anisotropy $\xi_r$ yields consistent values for both. Furthermore, the use of dynamical rather than quenched gauge configurations necessitates a simultaneous tuning of the two parameters $\xi_q$ and $\xi_g$.

\section{Simulation Details}

The quark action used in this study is fine-Wilson coarse-Hamber-Wu \cite{tunedpaper} with stout-link smearing \cite{smear}. This action has been designed for anisotropic lattices and has leading discretisation errors of $O(a_tM_q)$ at tree-level. \\
The action is given by $S_Q = \overline{\psi} M \psi$, where
\begin{eqnarray}
M\psi(x) & = &  \frac {1} {a_{t}} \{ (\mu_{r}ma_{t} + \frac{18s} {\xi_q}
 + r -\frac{1} {2u_{t}}[(r - \gamma_{0}) U_{t}(x) \psi(x + \hat{t})
 + (r + \gamma_{0})U^{\dagger}_{t}(x - \hat{t})\psi(x - \hat{t})] 
\nonumber\\
& & - \frac{1} {\xi_q} \sum_{i} [ \frac{1} {u_{s}} (4s - \frac{2} {3} \mu_{r}\gamma_{i}) U_{i}(x) \psi(x + \hat{\imath}) + \frac{1} {u_{s}} (4s + \frac{2} {3} \mu_{r} 
\gamma_{i})U^{\dagger}_{i} (x - \hat{\imath}) 
\psi(x - \hat{\imath}) \nonumber \\
& &- \frac{1} { u^{2}_{s}} (s - \frac{1} {12} \mu_{r} \gamma_{i} )
U_{i}(x) U_{i}(x + \hat{\imath}) \psi( x + 2 \hat{\imath} ) \nonumber \\
& &-\frac {1} { u^{2}_{s} } (s + \frac{1} {12} \mu_{r} \gamma_{i})
U^{\dagger}_{i} ( x - \hat{\imath} ) U^{\dagger}_{i} ( x - 2 \hat{\imath}) 
\psi( x - 2 \hat{\imath}) 
       ]  \} .
\end{eqnarray}
with $\mu_r = r = 1$, $s = \frac{1}{8}$ as in Ref.\cite{tunedpaper}

The tuning for the quenched case has previously been discussed in Ref. \cite{tunedpaper}. It was found that the tuned point had small mass dependence for a large range of quark masses from below strange to above charm. 

The gauge action used in these simulations is the two-plaquette Symanzik-improved action \cite{gluon_action}
\begin{eqnarray}
S_{G} = \frac{\beta} {\xi_{g}}
\left \{ \frac{5(1+\omega) } { 3 u^{4}_{s} } \Omega_{s}
 - \frac{5 \omega} {3 u^{8}_{s}} \Omega^{(2t)}_{s} 
- \frac{1} {12 u^{6}_{s}} \Omega^{(R)}_{s}
\right \}  + \beta \xi_{g}
\left \{
\frac{4} {3 u^{2}_{s} u^{2}_{t}} \Omega_{t}
- \frac{1} {12 u^{4}_{s} u^{2}_{t}} \Omega^{(R)}_{t}
\right \}.
\end{eqnarray}
where $\Omega_s$ and $\Omega_t$ are spatial and temporal plaquettes. $\Omega_s^R$ and $\Omega_t^R$ are $2 \times 1$ rectangles in the $(i,j)$ and $(i,t)$ planes respectively. $\Omega_s^{2t}$ is constructed from two spatial plaquettes separated by a single temporal link. 
This action has \begin{math} O(a^{4}_{s},a^{2}_{t}, \alpha_{s}a^{2}_{s}) \end{math}
discretisation errors.

\begin{table}
\begin{center}
\begin{tabular}{|c|c|}
\hline
\# gauge configurations & 250 \\
Volume & $8^3\times 48$ \\
$a_s$              & 0.2fm                    \\
Target Anisotropy $\xi_r=a_s/a_t$      & 6 \\
$a_tm_q$           & -0.057\\
\hline
\end{tabular}
\caption{Table of run parameters.}
\label{tab:run}
\end{center}
\end{table}

We used an $8^3 \times 48$ anisotropic lattice with lattice spacing $a_s = 0.2\mathrm{fm}$, and set a target anisotropy of 6. A set of 250 stout-link background gauge configurations was generated for each set of input parameters ($\xi_g$,$\xi_q$). Ten independent Markov chains were used. Approximately 5000 cpu hours were needed in order to generate each set of configurations. In order to increase statistics, 1000 bootstrapped sets of configurations were taken and analysis was done on these bootstrapped sets. All measurements were made using point propagators. Calculation of the correlation functions and analysis was negligible in comparison to the time needed to generate the gauge configurations ($\sim O$(30 hours)). 

\section{Tuning Method}
Unlike the quenched case \cite{tunedpaper} it is not possible to simply fix $\xi_g$ and then tune $\xi_q$ to a consistent value, since changing $\xi_q$ will affect the measurement of $\xi_g$. Explicitly, changing the value of $\xi_q$ necessitates  a regeneration of the background fields with the new value of $\xi_q$ which in turn will change the measured anisotropy $\xi_g$ of the background fields. The solution to this problem is a simultaneous two-dimensional tuning procedure. 

\begin{figure}
\includegraphics[scale=1.2]{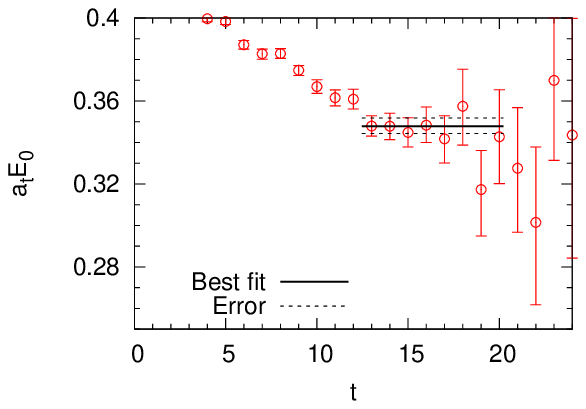}
\includegraphics[scale=1.2]{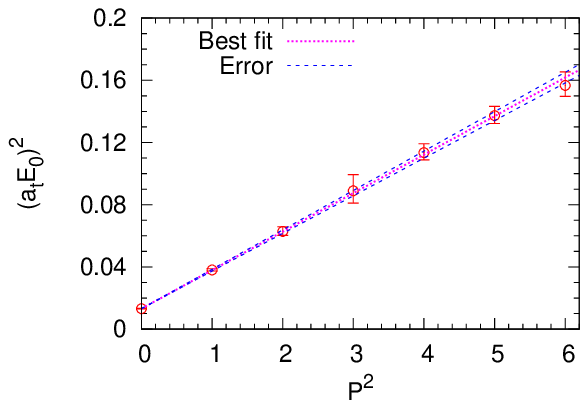}
\caption[Effective mass plot and Dispersion Relation]{The left plot shows the effective mass for degenerate quark mass $a_tm_q=-0.057$ momentum $n^2$ = 4. The fit range is [14:20] with a $\chi^2 /N_{dof} = 0.81$. The right plot is a dispersion relation with a $\chi^2 /N_{dof} = 0.14$. Both come from simulation parameters at point 1 on Fig. {\protect \ref{fig:all}}.}
\label{fig:eff}
\end{figure}

A linear dependence on the parameters $\xi_g$ and $\xi_g$ was assumed for a small region. Three initial sets of configurations were generated and the renormalised anisotropy was determined.
To determine the quark anisotropy for each simulation, the ground state energy, $E_0$, was determined for momenta with $n^2 = \{0,1,2,3,4,5,6\}$, $p_n = \frac{2 \pi n}{L a_s}$, from a single exponential $\chi^2$ minimisation fit to correlation functions. These values were then used to generate an energy momentum dispersion relation. The output quark anisotropy can be easily determined since it is inversely proportional to the square root of the slope of the dispersion relation. A sample effective mass plot and dispersion relation can be seen in Fig.~\ref{fig:eff}.
We employed the sideways potential\cite{sp1} method to tune the output gluon anisotropy. In this method a coarse direction on the lattice is chosen to be time. There are then both fine and coarse spatial directions. The static inter-quark potential is measured for separation both in the coarse and fine directions. The demand that the two measurements yield the same function of physical distance, $V_s(x) = V_t(\frac{t}{\xi})$, determines the renormalised anisotropy.

 Planes were defined for both output values of $\xi_g$ and $\xi_q$ i.e. values $\alpha,\beta,\gamma$ were found to satisfy $\xi_r = \alpha \xi_g + \beta \xi_q + \gamma$ for the renormalised anisotropy $\xi_r$ measured for each input $(\xi_q,\xi_g)$. The intersection of these planes with the required output value gave an intersection point. As bootstrapping was used it was necessary to ensure that each measurement of $\xi_g$ and $\xi_q$  used to compute an individual intersection point was computed from the same bootstrapped set. 
For the case of more than three sets of configurations, a plane was defined using a constrained-$\chi^2$ fit.
The target anisotropy was 6. One important point however is that in order to restore Lorentz symmetry the values $\xi_q$ and $\xi_g$ need only be consistent and the actual value they are tuned to is irrelevant.

\section{Results}
\begin{figure}
\includegraphics[scale=1.2]{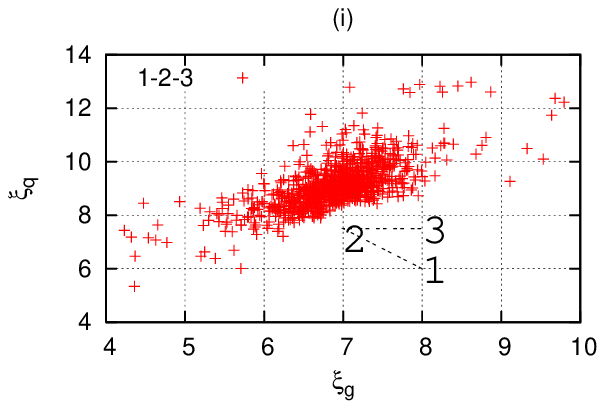}
\includegraphics[scale=1.2]{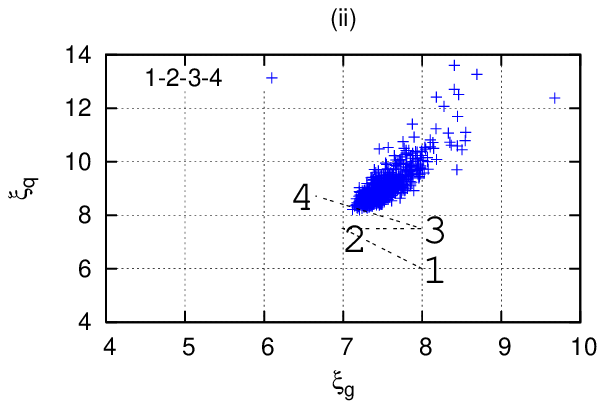}

\includegraphics[scale=1.2]{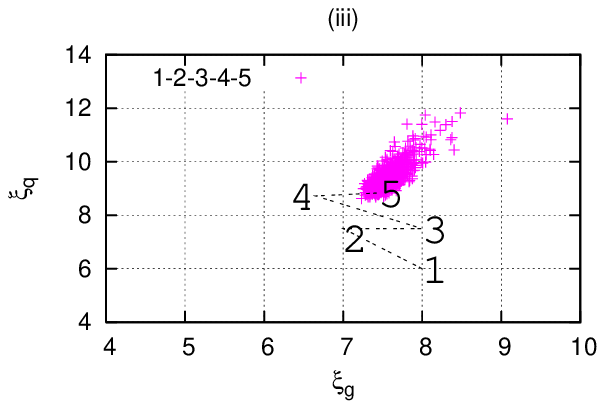}
\includegraphics[scale=1.2]{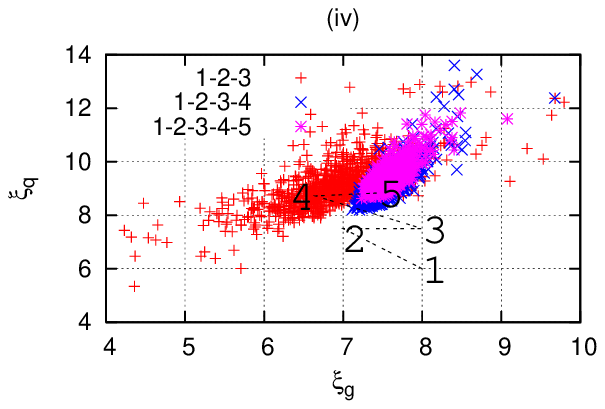}
\caption[Evolution of tuning procedure]{This figure shows the progression of the tuning procedure. (i) shows the scatterplot for a fit to the first 3 sets of configurations. (ii) shows a $\chi^2$ fit for the first 4. (iii) shows a $\chi^2$ fit for 5 sets. (iv) right shows all 3 scatterplots together. Each point on a scatterplot shows an estimate of a tuned point from different bootstrapped samples.}
\label{fig:all}
\end{figure}

\begin{figure}
\includegraphics[scale=1.2]{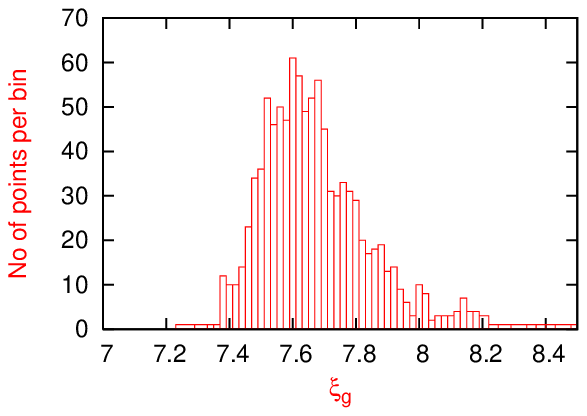}
\includegraphics[scale=1.2]{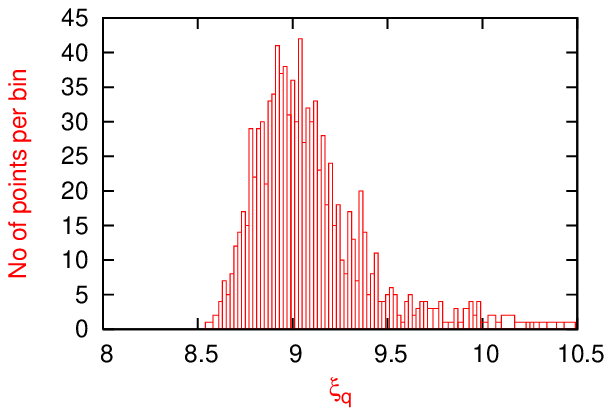}
\caption[Histogram of coordinates for scatterplot from Figure (\protect \ref{fig:all})]{Histogram of coordinates for 5-run scatterplot from Figure (\protect \ref{fig:all}). The plot on the left is the x-coordinate, $\xi_g$, plot on the right is for the y-coordinate $\xi_q$.}
\label{fig:histo}
\end{figure}

Figure~\ref{fig:all} shows the evolution of the tuning procedure. The first plot shows the intersection points obtained for the first three sets of configurations. The initial sets of ($\xi_g,\xi_q$) had been estimated to be close to the tuned values for the quenched case \cite{tunedpaper}. However the resulting scatterplot,(i) in Fig.~\ref{fig:all}, from the points of intersection lay outside the original triangle of points. Due to the large extrapolation, the error in central values was large. A fourth set of ($\xi_q,\xi_g$) was picked and a $\chi^2$ planar fit done for the four runs. This scatterplot is shown in (ii) Fig.~\ref{fig:all}. The result of this was a shifting of the central values and a large reduction in the error. A fifth set of configurations was generated and the process repeated. This resulted in a further reduction of the error in the central values. Figure~\ref{fig:histo} shows histograms where the coordinates of the points in this final scatterplot, (iii) Fig.~\ref{fig:all} were binned in bins of size $0.02$.
An analysis of the 1000 bootstrapped points found median values of $\xi_q = 9.02^{+0.32}_{-0.20} $ and  $\xi_g = 7.64^{+0.19}_{-0.13}$ using a  $68 \%$ confidence level.

\begin{table}
\begin{center}
\begin{tabular}{|c|c|c|c|c|}
\cline{2-5}
\multicolumn{1}{c}{}&\multicolumn{2}{|c|}{Input Parameters}&\multicolumn{2}{|c|}{Measured}\\
\hline
Point&$\xi_g$ &$\xi_q$& $\xi_q$ & $\frac{V_s(x)}{V_t(\frac{t}{\xi})}$ \\
\hline
\hline
 1&8.0 &6.0& 4.98 $\pm$ 0.06 &0.991 $\pm$ 0.003\\
\hline
 2&7.0 &7.5& 6.27$\pm$0.04& 0.986$\pm$ 0.003\\
\hline
 3&8.0 &7.5 &  5.18$\pm$ 0.006& 1.001 $\pm$0.003\\
\hline
 4&6.65 &8.72& 6.47$\pm$0.05& 0.985$\pm$0.005\\
\hline
 5&7.44 &8.83& 5.80$\pm$ 0.05& 0.995 $\pm$ 0.003\\
\hline
\end{tabular}
\end{center}
\caption{Table of measured output anisotropies measured at each of the run points in plot (iv) in Fig. {\protect \ref{fig:all}}.}
\label{tab:out}
\end{table}

\section{Conclusions and Outlook}
After generating five sets of configurations we have narrowed down the search for a tuned point to a small area. However, our choice of input parameters has meant that we are still extrapolating to estimate the tuned point. Our initial choice of parameters were too closely grouped. A better approach might have been to take a greater spread of points for our initial attempt at the tuning but the assumption of linear behaviour of $\xi_g $ and $\xi_q$ would have less validity over a larger region. 
This procedure will now be repeated using all-to-all propagators \cite{alan} and lattices with longer temporal extent which has been shown to improve determination of effective masses \cite{foley}. The final tuned point here will be used as a reference point. It is expected that there will be a small quark mass dependence on the tuned values for a large range of quark mass, similar to the quenched case \cite{tunedpaper}.

\section{Acknowledgements}
Richie Morrin is supported by the PRTLI IITAC grant


\begin{thebibliography}{99}
\bibitem{jonivar}R.~Morrin, A~ \'O Cais, M.B~Oktay, M.J.Peardon, J.I Skullerud, G. Aarts, C.R. Allton, \emph{Charmonium spectral functions in $N_f=2$ QCD}, PoS {\bf (LAT2005)} 176.
\bibitem{tunedpaper}J.~Foley et al., \emph{A non-perturbative study of the action parameters for anisotropic-lattice quarks}, hep-lat/0405030.
\bibitem{gluon_action} C. Morningstar, M.~J.~Peardon, \emph{The glueball spectrum from novel improved actions}, Nucl. Phys Proc Suppl, {\bf 83} (2000) 887-889 [hep-lat/9911003].
\bibitem{smear}C.~Morningstar and M.~J.~Peardon, \emph{Analytical smearing of SU(3) link variables in lattice QCD}, Phys. Rev. {\bf D69} (2004)054501 [heplat/0311018].
\bibitem{sp1}I.T~Drummond, R.R~Horgan, H.~Shanahan, M.~J.~Peardon, \emph{Measuring the aspect ratio renormalisation of anisotropic-lattice gluons}, hep-lat/0003019.
\bibitem{alan}J.~Foley et al., \emph{Practical all-to-all propagators for lattice QCD}, hep-lat/0505023.
\bibitem{foley}J.~Foley et al., \emph{ Static-light hadrons on a dynamical anisotropic lattice}, PoS {\bf(LAT2005)} 213.
\end{thebibliography}
\end{document}